\documentclass[%
 reprint,
superscriptaddress,
 amsmath,amssymb,
 aps,pra
]{revtex4-2}

\usepackage{graphicx}
\usepackage{dcolumn}
\usepackage{bm}
\usepackage[colorlinks=true,linkcolor=blue,urlcolor=blue,citecolor=blue]{hyperref}

\usepackage{subfigure}
\usepackage{xcolor}
\usepackage{amsmath}
\usepackage{physics}
\usepackage{float}

\usepackage[english]{babel}
\begin{document}

\preprint{APS/123-QED}
\title{Spatially resolved spin angular momentum mediated by spin-orbit interaction in tightly focused spinless vector beams in optical tweezers}

 \author{Ram Nandan Kumar}
 \email{rnk17ip025@iiserkol.ac.in}
 \affiliation{Department of Physical Sciences, Indian Institute of Science Education and Research Kolkata, Mohanpur-741246, West Bengal, India}

 \author{Sauvik Roy}
 \affiliation{Department of Physical Sciences, Indian Institute of Science Education and Research Kolkata, Mohanpur-741246, West Bengal, India}

\author{Subhasish Dutta Gupta}
\email{sdghyderabad@gmail.com}
\affiliation{Department of Physical Sciences, Indian Institute of Science Education and Research Kolkata, Mohanpur-741246, West Bengal, India}
\affiliation{Tata Institute of Fundamental Research, Hyderabad, Telangana 500046, India}
\affiliation{Department of Physics, Indian Institute of Technology, Jodhpur 342030, India}

\author{Nirmalya Ghosh}
\email{nghosh@iiserkol.ac.in}
\affiliation{Department of Physical Sciences, Indian Institute of Science Education and Research Kolkata, Mohanpur-741246, West Bengal, India}

\author{Ayan Banerjee}
 \email{ayan@iiserkol.ac.in}
 \affiliation{Department of Physical Sciences, Indian Institute of Science Education and Research Kolkata, Mohanpur-741246, West Bengal, India}

\date{\today}
\date{\today}

\begin{abstract}

We demonstrate an effective and optimal strategy for generating spatially resolved longitudinal spin angular momentum (LSAM) in optical tweezers by tightly focusing first-order azimuthally radially polarized (ARP) vector beams with zero intrinsic angular momentum into a refractive index (RI) stratified medium. The stratified medium gives rise to a spherically aberrated intensity profile near the focal region of the optical tweezers, with off-axis intensity lobes in the radial direction possessing opposite LSAM (helicities corresponding to $\sigma = +1$ and -1) compared to the beam centre. We trap mesoscopic birefringent particles in an off-axis intensity lobe as well as at the beam center by modifying the trapping plane, and observe particles spinning in opposite directions depending on their location. The direction of rotation depends on particle size with large particles spinning either clockwise (CW) or anticlockwise (ACW) depending on the direction of spirality of the polarization of the ARP vector beam after tight focusing, while smaller particles spin in both directions depending on their spatial location. Numerical simulations support our experimental observations. Our results introduce new avenues in spin-orbit optomechanics to facilitate novel yet straightforward avenues for exotic and complex particle manipulation in optical tweezers.

\end{abstract}


\maketitle

\section{Introduction}

Energy and momentum are the two fundamental dynamical quantities of light, each manifesting through the respective conservation laws \cite{bliokh2014magnetoelectric,bliokh2009geometrodynamics}. Among these, momentum has attracted considerable attention in elucidating fundamental photonic interactions. The components of momentum — linear momentum, spin angular momentum (SAM), orbital angular momentum (OAM), and their inter-conversions, known as the spin-orbit interaction (SOI) of light \cite{bliokh2013dual,andrews2012angular,bliokh2015spin}— play a crucial role in all light-matter interactions. The SOI of light is manifested in numerous optical elements encompassing the in-plane and out-of-plane  Goos-Hänchen (GH) and Imbert-Fedorov (IF) shifts \cite{bliokh2015transverse}, akin to the spin-Hall effect of bounded beams undergoing reflection or transmission from planer interfaces \cite{Int_10,Int_12, Int_11}, spin-dependent optical vortex generation in tight focusing \cite{fu2019spin} as well as in the subwavelength Epsilon-Near-Zero (ENZ) slabs \cite{eismann2022enhanced}, spin-hall effects in scattering  \cite{bliokh2015spin} and tight focusing \cite{kumar2022probing}, etc. Some of the recent SOI-driven phenomena include unusual transverse spin angular momentum (TSAM) both in evanescent wave  \cite{aiello2015transverse,neugebauer2015measuring} and tight focusing of circularly polarized (CP) light \cite{stilgoe2022controlled,pal2020direct,chen2022spin, shao2018spin,saha2018transversespin}, simultaneous spin-dependent directional guiding \cite{bliokh2015spin}, and wavevector-dependent spin acquisition (spin-direction-spin coupling) in plasmonic crystals \cite{nayak2023spin}, rotational spin-hall effect in structured Gaussian beams \cite{kumar2022probing}, etc.

Apart from these manifestations, several other crucial aspects emerged in tightly focused fields, particularly in particle manipulation using optical tweezers \cite{kumar2024inhomogeneous, youngworth2000focusing,yang2021non,yang2021optical}. The controllable spinning of microparticles employing orbital-Hall effect (OHE) \cite{li2015observation, Int_10, Int_11, Int_12, Int_13}, spin-momentum locking due to tight focusing of CP light \cite{bliokh2015spin,bliokh2014extraordinary}, orbital motion around the beam axis due to transfer of OAM  \cite{he1995direct} as well as due to enhanced SOI in a stratified medium \cite{kumar2024inhomogeneous,kumar2024probing}, and Larmor-like concurrent precessional and partial orbital motion  \cite{Roy2024observation} are a few instances of SOI driven opto-mechanical effects. It is therefore quite clear that SOI can be efficiently utilized to facilitate complex light-driven motion of microparticles by precisely tailoring the focused field to meet specific requirements\cite{zeng2024tightly,chen2022spin}. The necessary field can be crafted by modifying the surrounding propagating mediums near the focus or by choosing beams with diverse field parameters, such as spin angular momentum (SAM), orbital angular momentum (OAM) polarization, and intensity distribution \cite{andrews2012angular,zhou2023polarization,he1995direct,o2014spin}. Among the various types of beams, cylindrical vector beams have attracted significant attention due to the cylindrical or axial symmetry around the beam axis \cite{zhan2009cylindrical,berry2009optical,beresna2011radially,kumar2022manipulation} and the inhomogeneous polarization across the beam cross-section \cite{forbes2019structured,forbes2021structured,zeng2018sharply}. These vector beams are higher-order solutions of the paraxial vector Helmholtz equation and can be represented as superpositions of Hermite-Gaussian (HG) or Laguerre-Gaussian (LG) modes\cite{gupta2015wave, Novotny2012,kumar2024rectangular}. For the last two decades, these vector beams have been employed in a wide variety of both fundamental and application-based research including multiplexing \cite{gregg2019enhanced}, quantum sensing \cite{degen2017quantum, pirandola2018advances}, quantum information \cite{bennett2000quantum, ladd2010quantum}, optical communication \cite{ren2016orbital}, particle trapping \cite{padgett2011tweezers,o2002intrinsic}, optical encryption \cite{liu2017single}, quantum memory \cite{nicolas2014quantum}, super-resolution microscopy \cite{torok2004use}, etc. 

Among numerous vector beams, radially and azimuthally polarized beams are the ones frequently encountered \cite{andrews2012angular,zhan2009cylindrical,peters2024controlling}. By combining these two beams, an azimuthally radially polarized (ARP) vector beam can be generated, exhibiting either clockwise (CW) or anticlockwise (ACW) spiral polarization distribution depending on the superposition type. As the constituent radial and azimuthal beams lack orbital or spin angular momentum, the resultant ARP vector beams also possess a net zero angular momentum, with individual momenta being zero separately \cite{xu2019azimuthal,veysi2016focused,chen2021engineering}. These ARP vector beams exhibit cylindrical symmetry about the propagation axis z, akin to their constituent beams \cite{zhan2009cylindrical,forbes2021structured}. Motivated by the unique polarization properties of the ARP vector beams, and given that the effects of SOI lead to generation of spin from input spinless beams after tight focusing, etc \cite{roy2014manifestations,kumar2024probing},  an interesting question to ask is whether ARP vector beams can be coupled into optical tweezers to generate intriguing  rotational or spin dynamics of trapped birefringent microparticles (which can exchange spin angular momentum with light). Particularly, could one actually engineer an optical trap with different spatial regions possessing different spin polarization, so that particles trapped in those regions would spin differently? This is what we study in this paper where we tightly focus ARP vector beams through a high numerical aperture (NA) objective lens into a refractive index (RI) stratified medium. Our study not only reveals that these `spin-less' beams are capable of driving rotational motion in birefringent micro-particles, but also showcases site and size-specific control of such spinning motions -- hitherto unfeasible or inaccessible for other types of beams to the best of our knowledge. Specifically, both clockwise and anticlockwise spinning motions depending on the size of the particles are achieved at different spatially separated regions. The underlying reason is found to stem from the distribution of the longitudinal spin angular momentum density (LSAM) caused by tight focusing into the RI-stratified medium depicting a distinctive form of the spin Hall effect. Indeed, the RI-stratified medium also broadens the permissible parameter space suitable for practical usage of the novel particle manipulation technique. Our experimental results are  substantiated by the simulations performed under the light of full vectorial Debye-Wolf diffraction theory for high NA focusing.

The structure of the paper is as follows: In Sec. II, experimental findings are described. In Sec III, the theoretical framework for the focused field of the incident ARP vector beams is outlined, which are then explained in terms of simulated field quantities in Sec IV. An overall summary of the study can be found in Sec V. 

\begin{figure*}
\includegraphics[width=\textwidth]{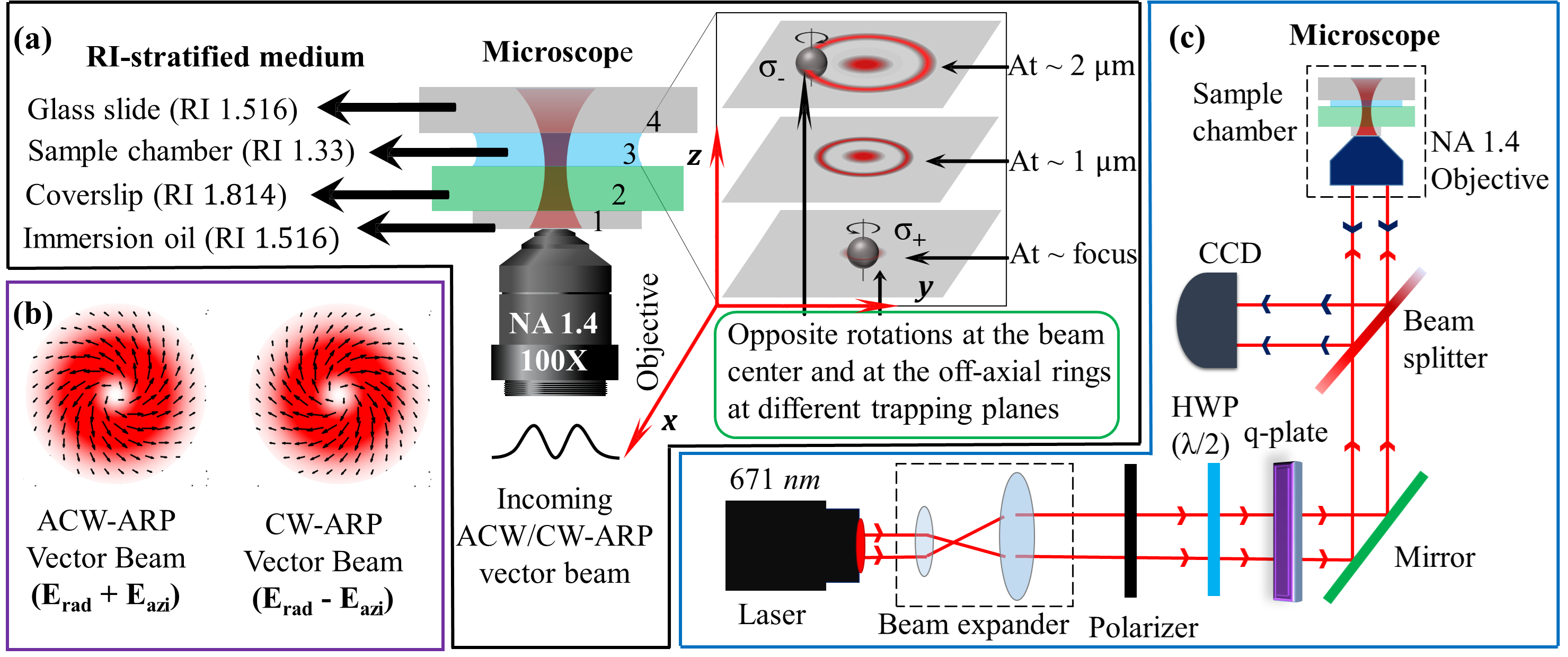}
\caption{Schematic of the experimental setup for unconventional optical tweezers utilizing a refractive index (RI) stratified medium, coupled with ARP vector beams. (a) Illustration of the RI stratified medium composed of four layers, with the third layer (sample chamber) where particles are trapped at on-axis and off-axis positions, showing particles rotating in opposite directions relative to each other. (b) Depiction of the doughnut intensity pattern and spiral polarization distribution of the input ACW-ARP and CW-ARP vector beams. (c) Ray diagram of the experimental setup, showing how ACW/CW-ARP vector-polarized light is coupled into the optical tweezers for probing the spatially resolved LSAM.}
\label{schematic}
\end{figure*}

\section{Experimental results}

In our experiments using optical tweezers, two aspects -- (a) generating the ARP vector beams and (b) a stratified medium featuring varied refractive index (RI) layers to control the nature of the SOI -- are of primary importance. The schematic and details of our optical tweezers setup are shown in Figs.~\ref{schematic} (a)-(c). A polarizer, a half-wave plate (HWP), and a zero-order vortex plate (q-plate) designed for a wavelength of 671 nm (arranged in the sequence shown in Fig.~\ref{schematic}(c)) were utilized to convert the incident Gaussian laser beam of wavelength 671 nm (Lasever, 350 mW) into ARP vector beams. The polarizer's axis was fixed at $0^\circ$, and the fast axis of the half-wave plate was set at $22.5^\circ$ and $67.5^\circ$ with respect to the polarizer axis, so that $45^\circ$ ($E^{in}=\frac{E_0}{\sqrt{2}}\begin{bmatrix}1 & 1\\\end{bmatrix}^T$) and $135^\circ$ ($E^{in}=\frac{E_0}{\sqrt{2}}\begin{bmatrix}1 & -1\\\end{bmatrix}^T$) linearly polarized light were produced, respectively. The q-plate's axis was then aligned at $0^\circ$ (horizontal direction) to convert the $45^\circ$ linearly polarized beam into an ACW-ARP vector beam ($\mathbf{E}_{\text{rad}} +\mathbf{E}_{\text{azi}}$ ;~$E^{ACW}_{ARP}=\frac{E^{\prime}_0}{\sqrt{2}}\begin{bmatrix}\cos \phi + \sin \phi ~ & \sin \phi - \cos \phi\\\end{bmatrix}^T$) and the $135^\circ$ linearly polarized beam into a CW-ARP vector beam ($\mathbf{E}_{\text{rad}} -\mathbf{E}_{\text{azi}}$ ;~$E^{CW}_{ARP}=\frac{E^{\prime}_0}{\sqrt{2}}\begin{bmatrix}\cos \phi - \sin \phi ~ & \sin \phi + \cos \phi\\\end{bmatrix}^T$). Where $E_0$ and $E^{\prime}_0$ represent the Gaussian and doughnut-shaped amplitude profiles, respectively. The resulting spiral polarization distributions of the ACW-ARP and CW-ARP vector beams, plotted over the doughnut-shaped intensity pattern before focusing, are shown in Fig.~\ref{schematic}(b).

These beams were then guided into the back-port of the inverted microscope (Carl Zeiss Axioert.A1) equipped with an objective lens with NA 1.4. (b) The RI-stratified medium, comprising four layers—immersion oil (RI 1.516), a coverslip (RI 1.814), the sample dispersed in deionized (DI) water (RI 1.33), and a top glass slide (RI 1.516)—constitutes our sample-mounting 1D inhomogeneous structure, as shown in Fig.~\ref{schematic}(a). While it is expected to use a coverslip with RI matching that of the immersion oil, employing one with a mismatched RI is particularly efficacious for our purposes due to the increased RI contrast in the stratified medium; we refer to this scenario as the `mismatched case' while the former is called the `matched case'\cite{kumar2024probing,kumar2024inhomogeneous,roy2013controlled}. 

The experimental results showcased involved a coverslip with a RI of $1.814$, deliberately chosen for its substantial mismatch with that of the immersion oil. It is noteworthy that as the immersion oil (first layer) remains the same throughout the experiment, the refractive index of the coverslip (second layer) becomes pivotal in altering the behavior of light within the sample chamber (third layer), potentially increasing the capabilities of micro-particle manipulation as desired. The effects of the focused field within the sample chamber were visualized using spin-responsive micro-particles derived from an anisotropic liquid crystal called 5CB (4'-Pentyl-4-biphenylcarbonitrile from Sigma-Aldrich)\cite{lv2016photocontrol,de1993physics}. A dispersion of these micro-particles was prepared by mixing $5\mu l$ of 5CB, $100\mu l$ of polyvinyl alcohol (PVA), and $200ml$ of deionized water in a micro-centrifuge tube, vigorously shaken for several minutes. The particles/droplets had radii ranging from $0.5\mu m$ to $10\mu m$ and predominantly exhibited bipolar behavior, as evident from the intensity patterns observed across the cross-section of the particles due to the cross-polarization technique used for imaging \cite{muvsevivc2017liquid,brasselet2009optical}. 

\begin{figure*}
\includegraphics[width=\textwidth]{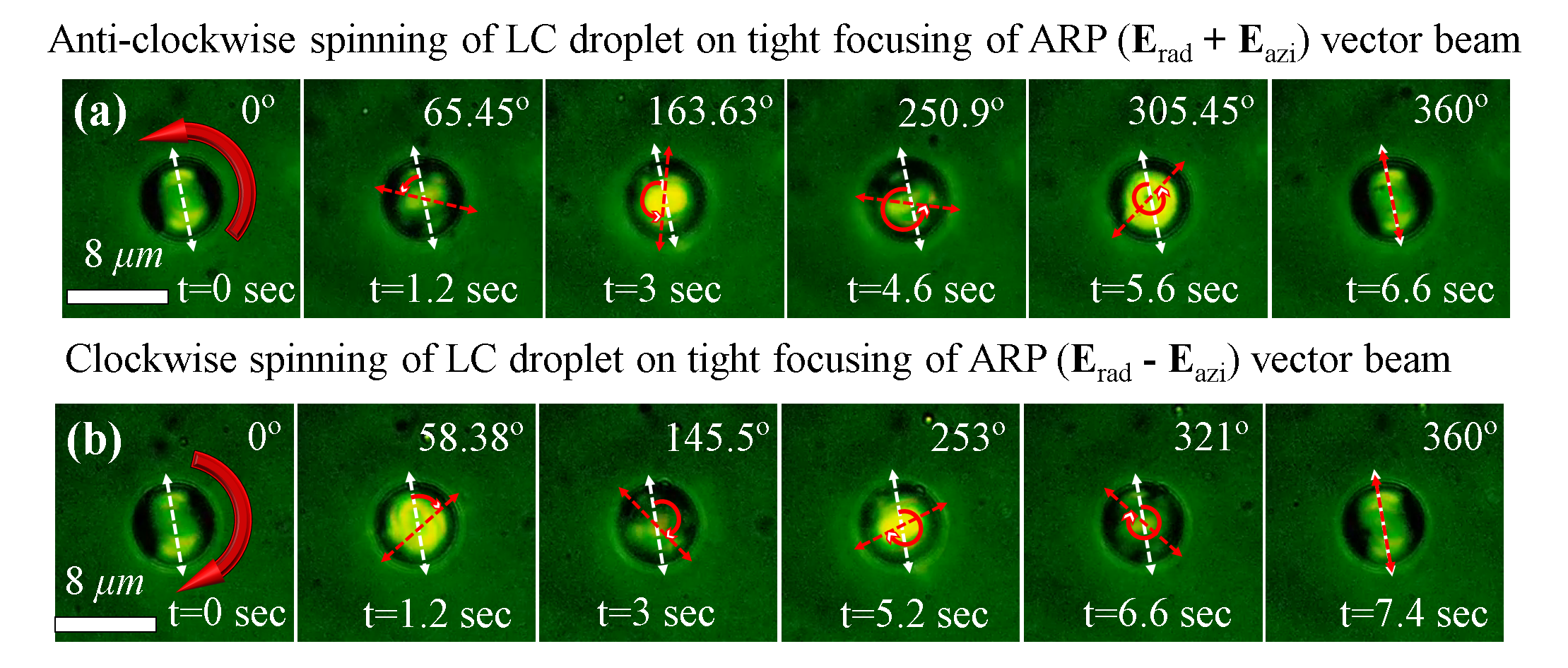}
\caption{Timelapse frames displaying the (a) anticlockwise and (b) clockwise rotation of the liquid crystal particles trapped at the beam center for the incident ACW-ARP ($\mathbf{E}_{\text{rad}} +\mathbf{E}_{\text{azi}}$) and CW-ARP ($\mathbf{E}_{\text{rad}} -\mathbf{E}_{\text{azi}}$) vector beams respectively. The particles being relatively bigger (diameter $\sim 7-8~\mu$m) follow the inherent spiral field orientations of the respective beams. White arrows serve as the reference (drawn at the onset of rotation) and the red arrows depict the instantaneous orientations.
}
\label{big_particle}
\end{figure*}
\begin{center}
\begin{figure}
\includegraphics[width=0.5\textwidth]{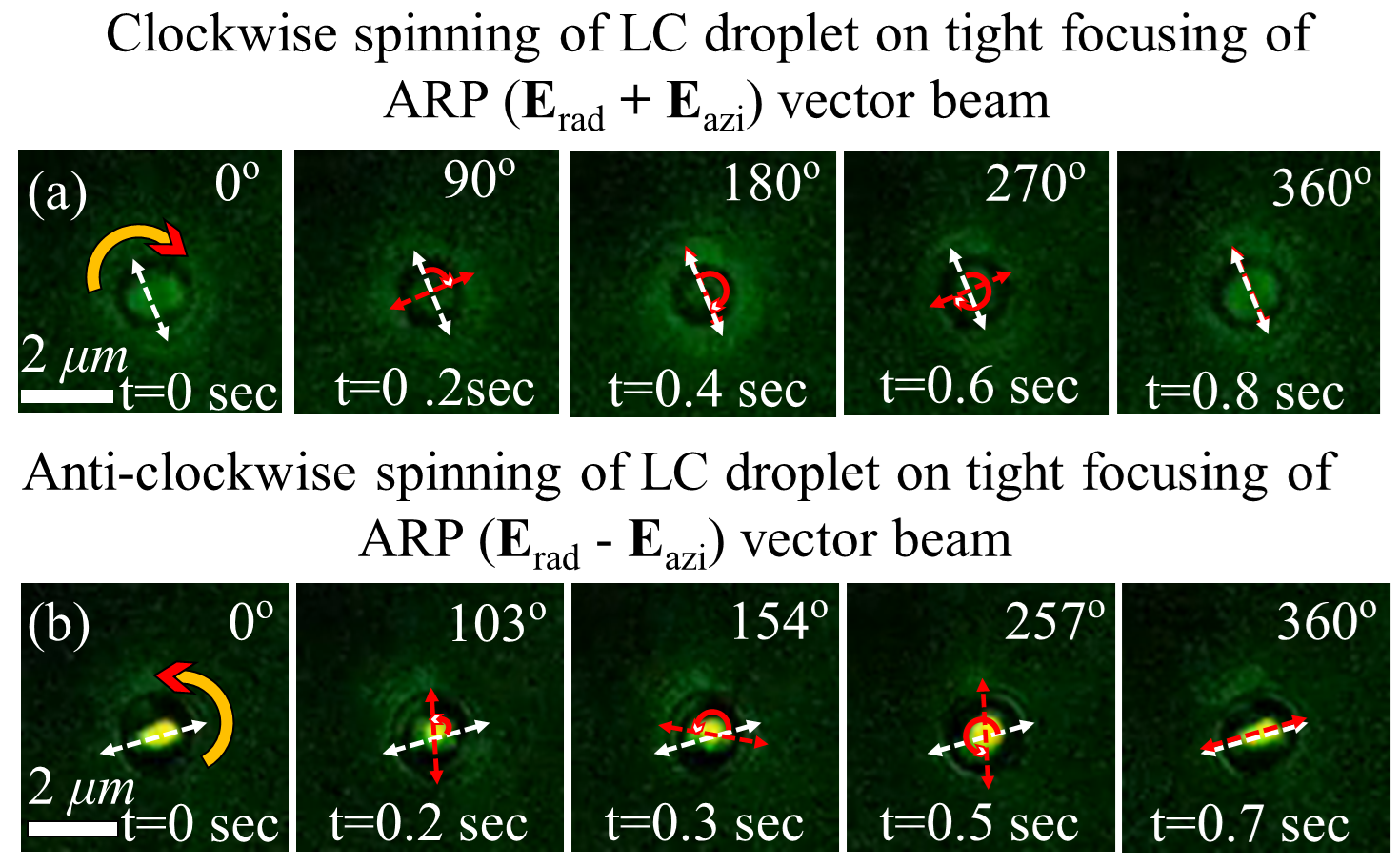}
\caption{Timelapse frames displaying the (a) clockwise and (b) anticlockwise rotation of the liquid crystal particles trapped at off-center locations for the incident ACW-ARP ($\mathbf{E}_{\text{rad}} +\mathbf{E}_{\text{azi}}$) and CW-ARP ($\mathbf{E}_{\text{rad}} -\mathbf{E}_{\text{azi}}$) vector beams respectively. Both particles (diameter $\sim 2\mu m$) spin in a direction opposite to the inherent spiral field orientations of the respective beams. White arrows serve as the reference (at the onset of rotation) and the red arrows depict the instantaneous orientations. }
\label{small_particle}
\end{figure}
\end{center}

The unconventional trapping mechanism utilizing the RI stratified medium focuses the field into an off-axis, spherically aberrated, ring-shaped region while also concentrating a significant amount of energy at the beam center (as detailed in Sec. IV). As a result, both the off-axis intensity annular ring and the on-axis trap center are suitable for trapping microparticles. Depending on the position where the liquid crystal particles are trapped, we can probe the spatially resolved both ($\sigma_{-}=-1$ and $\sigma_{+}=+1$) helicity of LSAM, which we describe in detail below.

The schematic detailing the spatially resolved \(\sigma_{-}\) and \(\sigma_{+}\) helicity of LSAM is shown in Fig.~\ref{schematic}(a), while the experimental results are presented in Figs.~\ref{big_particle}(a)-(b) and Figs.~\ref{small_particle}(a)-(b).
First, particles size, typically ranging from $3-4~\mu$m in radius and trapped at the beam center exhibited anti-clockwise rotational motion for the ACW-ARP vector beam, and clockwise rotation for the CW-ARP vector beam, respectively, as comprehended from the rotation of the crossed-patterns on the particles (see Video 1 and 2 in the online Supplementary Information). Expectedly, at the beam center, the observed motion aligned with the inherent ACW and CW spiral patterns of the electric fields (or polarization) at the respective beams' cross-sections (see Fig. \ref{big_particle} (a)-(b)). Second, in the case of smaller particles, typically with radii ranging from $0.5-1~\mu m$ trapped in an off-axial location, unexpected events were observed: clockwise rotations were noted for the ACW-ARP vector beam, while anticlockwise rotations were observed for the CW-ARP vector beam (see Video 3 and 4 in the online Supplementary Information). Thus, for the same incident ARP vector beam, either ACW-ARP or CW-ARP, a reversal in directions was observed when comparing the rotations of the larger particles (at the axial location) with that of the smaller particles (at the off-axial location).  The time-lapse frames for two different-sized particles, one with diameter $\sim 7-8~\mu$m  and another $ \sim 2~\mu$m, display these size-dependent opposite rotations in Figures \ref{big_particle} and \ref{small_particle}, respectively. Indeed, for small-sized particles \(\sim 1-2 \, \mu m\) in diameter, we could switch the rotation from clockwise to anti-clockwise by adjusting the focus of the microscope objective lens. As observed in Video 5, the particle initially spins in a clockwise direction at the center of the beam. However, as we change the focus, the particle shifts its position from the on-axis (beam center) to an off-axis position in the trap (due to the radial intensity gradient). It starts rotating in an anti-clockwise direction because the helicity of the LSAM at the off-axis position is opposite to that at the beam center for an input CW-ARP vector beam. Notably, the rotation frequency of the off-axis trapped particle is approximately four times less than that of the on-axis trapped same-size particle, which is expected due to the highly divergent nature of the beam. Such focusing-dependent change of the rotation direction (clockwise or anti-clockwise) offers considerable flexibility in controlling the spin of particles by adjusting the focusing objective lens during an experiment.

Additionally, some particles were observed to periodically exhibit subtle focusing and defocusing effects while rotating in either clockwise or anticlockwise directions. Typically, the clockwise/anticlockwise rotations stem from the longitudinal SAM, while the focusing and defocusing effects result from the transverse spinning of the particles due to the transverse SAM (see Video 3  in the online Supplementary Information). Nevertheless, this motion was observed to be significantly less frequent compared to the size-dependent rotational motion. Indeed, such size-dependent opposite rotational motion exhibited by bipolar particles, when trapped by a single spinless ARP vector beams at the beam center (on-axis) and off-axis intensity annular ring, stand as the key experimental discovery of this investigation. The ability of ARP vector beams to impart clockwise and counterclockwise torques prompt a close examination of the intensity and spin distribution near the focal region of the tightly focused polarized light, as we now proceed to carry out using the Debye-Wolf integrals described below. 

\begin{figure*}
\includegraphics[width=\textwidth]{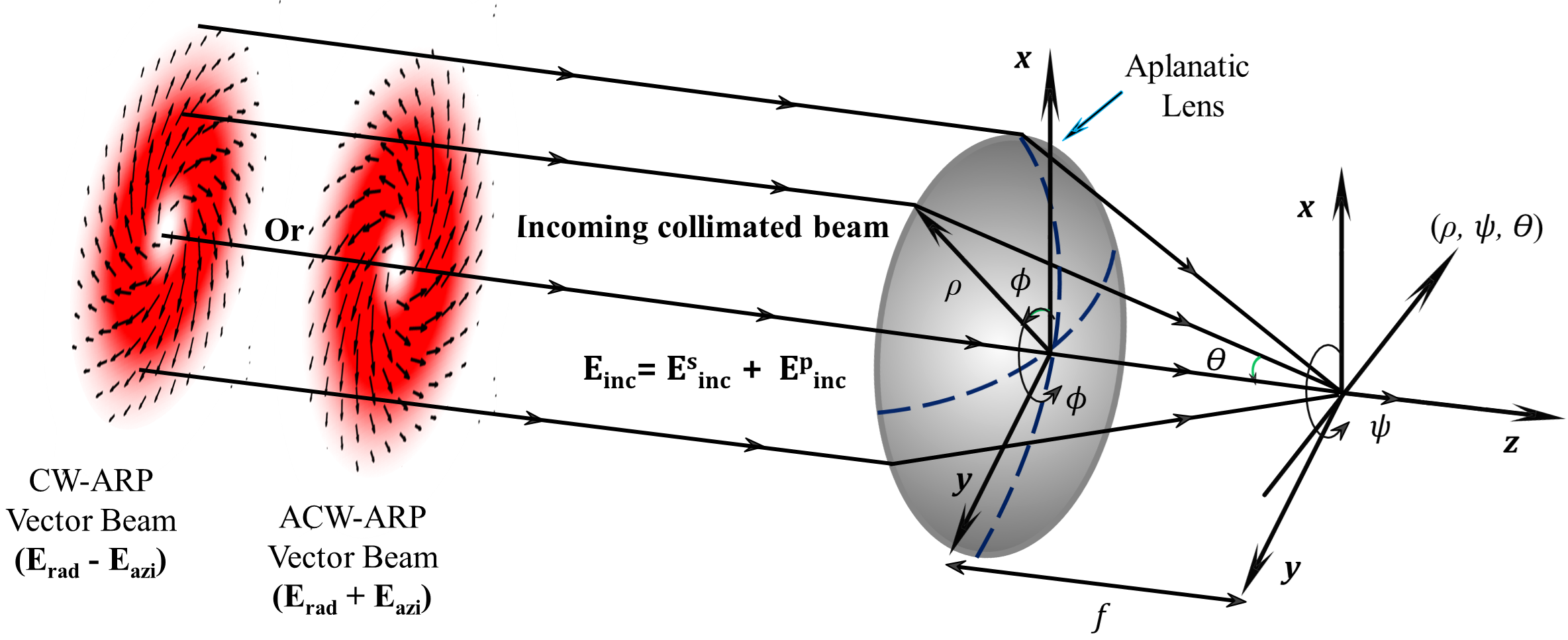}
\caption{Schematic of the geometric representation of the coordinate transformation from cylindrical to spherical coordinates through an aplanatic lens.}
\label{TF_sch}
\end{figure*}

\section{Theoretical Calculations}

Details of our theoretical formalism are provided in the online Supplementary Information - here we provide the basic outline, As mentioned previously, radially, azimuthally, and ARP vector beams are solutions of the paraxial vector Helmholtz equation \(\nabla \times \nabla \times \vec{E} - k^2 \vec{E} = 0\), which can be expressed as follows \cite{zhan2009cylindrical,xu2019azimuthal}:

\begin{equation}
\vec{E}_{rad/azi}(\rho, z)=(E_{0}/w_0)~ \rho \exp \left(-\frac{\rho^2}{w^2}\right) {\vec{e}}_i, \quad i=r, \phi
\label{Eq1}
\end{equation}

Here, \(i = \rho\) and \(i = \phi\) represent the radial (\(\vec{e}_\rho = \begin{bmatrix}\cos{\phi} & \sin{\phi} \end{bmatrix}^T\)) and azimuthal (\(\vec{e}_\phi = \begin{bmatrix}\sin{\phi} & -\cos{\phi} \end{bmatrix}^T\)) polarizations of the beam, respectively. The ARP vector beam before focusing can be expressed as a linear superposition of radially and azimuthally polarized vector beams:

\begin{equation}
\begin{aligned}
\vec{E}^{ACW/CW}_{ARP}(\rho,\phi, z)=(E_{0}/w_0)~ \rho \exp \left(-\frac{\rho^2}{w^2}\right)\\ \times\left(\begin{array}{c}\cos \phi \pm \sin \phi \\ \sin \phi \mp \cos \phi\end{array}\right), 
\label{Eq2}
\end{aligned}
\end{equation}

Here, \(E_{0}\) is the overall amplitude factor, \(w\) is the size parameter of the beam, and \(\rho^2 = x^2 + y^2\) is the radial distance from the central wave vector \(k\) (or beam center). The terms \(E^{ACW}_{ARP} = E_{rad} + E_{azi}\) and \(E^{CW}_{ARP} = E_{rad} - E_{azi}\) describe the anti-clockwise and clockwise spiral polarization directions of the ARP vector beam, respectively. It is important to note that only the first-order vector beams are considered throughout this study. The expressions in Equations (\ref{Eq1}) and (\ref{Eq2}) follow the paraxial approximations and can be directly integrated with the angular spectrum-based Debye-Wolf formalism for tight focusing by a high NA objective lens \cite{richards1959electromagnetic,Novotny2012}.

The Debye-Wolf formalism portrays the refracted, highly nonparaxial spherical wavefront emanating from an aplanatic lens into infinite plane waves commonly depicted as spatial harmonics. The transformations of the electric and magnetic field components are mimicked by the lens's transfer function, which relies on the decomposition of the incoming fields into TE (s-polarization) and TM components (p-polarization) at the lens's surface, as shown in Fig.~\ref{TF_sch} \cite{Novotny2012,roy2013controlled}. Subsequent modifications of the field components due to the interfaces formed by the stratification of different layers in an experimental environment are also incorporated in a similar decomposition manner at each interface. Hence, the lens itself and Fresnel's transmission and reflection coefficients are the two key ingredients in desirably tailoring the focus field. The resultant field is finally obtained by the linear superposition of the TE and TM components of all the plane waves, thus retaining the full vectorial details of the field in the image plane. According to this formalism, the time-independent monochromatic focused electric field is given by the angular spectrum integral \cite{Novotny2012,richards1959electromagnetic}
\begin{equation}
\begin{aligned}
    \textbf{E}(\rho,\psi,z) =
\int_{0}^{\theta_{max}}\int_{0}^{2\pi}
 E_{\infty}(\theta,\phi)e^{i k z \cos{\theta}}\\
 \times e^{i k\rho \sin{\theta} \cos{(\phi-\psi)}} \sin{\theta} d\theta d\phi
 \end{aligned}
 \label{angular}
\end{equation}

Here, we neglected the evanescent fields and considered only the far-field component, denoted as $E_{\infty}(\theta, \phi)$. The wave-vector in the medium is $k$, \( \theta_{\max} = \sin^{-1}(\text{NA} / n) \) is the maximum angle determined by the NA of the objective lens, while \( n \) denotes the refractive index of the medium. $E_{\infty}(\theta, \phi)$ for the ACW-ARP/CW-ARP vector beam is calculated using a coordinate transformation through the aplanatic lens (or the lens's transfer function)  as follows:

\begin{equation}
\begin{aligned}
E^{ARP}_{\infty}(\theta, \phi)=\left( E_0 f / \omega_0\right) ~f_\omega(\theta)~\sqrt{\frac{n_1}{n_2}}(\cos \theta)^{1 / 2}~\sin \theta~\\
 \times \left[\begin{array}{c}
t^p \cos \theta~\cos  \phi \pm t^s ~\sin  \phi \\
t^p \cos \theta~\sin  \phi \mp t^s ~\cos  \phi\\
-~t^p \sin \theta
\end{array}\right]
\end{aligned}
\label{9}
\end{equation}

Here, the function \(f_\omega(\theta)\) is the apodization function that arises when an aplanatic lens tightly focuses the beam. By substituting Eq.~\ref{9} into Eq.~\ref{angular}, and then evaluating the integration over \(\phi\), we obtain the equation for the focused electric field, which involves a single integration over the variable \(\theta\) \cite{Novotny2012, richards1959electromagnetic}. It is important to note that the magnetic field can be derived in a similar manner. For the injected ACW-ARP/CW-ARP vector beams, the focused electric and magnetic field components can be expressed in terms of the Debye–Wolf (or diffraction) integrals in Cartesian coordinates as follows:

\begin{equation}\label{electric}
\begin{aligned}
\left[\begin{array}{c}
E_x \\
E_y \\
E_z
\end{array}\right]_{\text{ARP}}^{\text {ACW/CW}} =\left[\begin{array}{c}
i\left(I_{1} \cos{\psi} \pm I_{2} \sin{\psi} \right) \\
i\left(I_{1} \sin{\psi} \mp I_{2} \cos{\psi} \right) \\
- I_{0}
\end{array}\right]
\end{aligned}
\end{equation}

\begin{equation}\label{magnetic}
\begin{aligned}
\left[\begin{array}{c}
H_x \\
H_y \\
H_z
\end{array}\right]_{\text{ARP}}^{\text {ACW/CW}}=\left[\begin{array}{c}
\pm i\left(I_{1} \cos{\psi} \mp I_{2} \sin{\psi} \right) \\
\pm i\left(I_{1} \sin{\psi} \pm I_{2} \cos{\psi} \right) \\
\mp  I_{0}
\end{array}\right]
\end{aligned}
\end{equation}

Where \( \Vec{E}_{\text{ARP}}^{\text{ACW/CW}} = \vec{E}_{\text{rad}} \pm \vec{E}_{\text{azi}} \) and \( \Vec{H}_{\text{ARP}}^{\text{ACW/CW}} = \vec{H}_{\text{rad}} \pm \vec{H}_{\text{azi}} \) are the electric and magnetic fields of the focused light, respectively. Note that the magnetic field is CW in nature for the input ACW-ARP vector beam; however, it is ACW in nature for the input CW-ARP vector beam, with equal and opposite non-zero z-components in both cases. The Debye-Wolf diffraction integrals for the transmitted and reflected waves, \( I_{0} = I_{0}^{t}(\rho) + I_{0}^{r}(\rho) \), \( I_{1} = I_{1}^{t}(\rho) + I_{1}^{r}(\rho) \) and \( I_{2} = I_{2}^{t}(\rho) + I_{2}^{r}(\rho) \), are determined by the polar angles of incidence (\(\theta\)) of the plane waves, and by Fresnel's transmission (\( t^s, t^p \)) and reflection (\( r^s, r^p \)) coefficients. The strength of the spin-orbit conversion is encapsulated by these integrals, which are given as \cite{Novotny2012,roy2013controlled,gupta2015wave,kumar2024probing}:

\begin{equation}
\scalebox{0.85}{$
\begin{aligned}
I_{1}^{t}(\rho)=& \int_0^{\theta_{\max }} f_\omega(\theta)\sqrt{\cos \theta} \sin ^{2} \theta \left(t^{p} \cos \theta \right) J_{1}\left(k \rho \sin \theta\right) e^{i k z \cos \theta} d \theta \\
 I_{2}^{t}(\rho)=& \int_0^{\theta_{\max }} f_\omega(\theta) \sqrt{\cos \theta} \sin ^{2} \theta~\left(t^{s}\right) J_{1}\left(k \rho \sin \theta\right) e^{i k z \cos \theta}  d \theta,\\
I_{0}^{t}(\rho)= & \int_0^{\theta_{\max }} f_\omega(\theta) \sqrt{\cos \theta} \sin ^{2} \theta \left(t^{p} \sin \theta\right) J_{0}\left(k \rho \sin \theta\right) e^{i k z \cos \theta} d \theta,\\
I_{1}^{r}(\rho)=& \int_0^{\theta_{\max }}- f_\omega(\theta)\sqrt{\cos \theta} \sin ^{2} \theta \left(r^{p} \cos \theta\right) J_{1}\left(k \rho \sin \theta\right) e^{-i k z \cos \theta} d \theta, \\
I_{2}^{r}(\rho)=& \int_0^{\theta_{\max }} f_\omega(\theta) \sqrt{\cos \theta} \sin ^{2} \theta \left(r^{s}\right) J_{1}\left(k \rho \sin \theta\right) e^{-i k z \cos \theta} d \theta,\\
I_{0}^{r}(\rho)=& \int_0^{\theta_{\max }} f_\omega(\theta) \sqrt{\cos \theta} \sin ^{2} \theta \left(r^{p} \sin \theta\right) J_{0}\left(k \rho \sin \theta\right) e^{-i k z \cos \theta} d \theta 
\end{aligned}
$}
\label{13}
\end{equation}

Where the superscripts \( t \) and \( r \) indicate the transmitted and reflected components, respectively. The functions \( J_0 \) and \( J_1 \) are the zero and first-order Bessel functions of the first kind, respectively. Due to the linearity of the system, the focused ARP vector beams in a nonmagnetic medium (i.e., \(\mu = \mu_{0}\)) are a linear superposition of the focused radially and azimuthally polarized vector beams. Following this, we calculate the spin angular momentum density, \(\mathbf{S}\), which is one of the key electromagnetic dynamical quantities and can be explicitly written as \cite{aiello2015transverse,neugebauer2015measuring,bliokh2014extraordinary}

\begin{equation}
    \mathbf{S} = \operatorname{Im}\left[\epsilon\left(\mathbf{E}^* \times \mathbf{E}\right)+\mu\left(\mathbf{H}^* \times \mathbf{H}\right)\right]
\end{equation}

The SAM density of the focused ACW-ARP/CW-ARP vector beam in Cartesian coordinates can be recast as

\begin{equation}
\begin{aligned}
& S_x^e= i(I_{1}^{\prime} \sin{\psi} \mp I_{2}^{\prime} \cos{\psi}) \\
& S_y^e=- i( I_{1}^{\prime}\cos{\psi} \pm  I_{2}^{\prime}\sin{\psi}) \\
& S_z^e=\mp (I_{1}^* I_{2}-I_{1} I_{2}^*)
\end{aligned}
\end{equation} 
Where, $I_{1}^{\prime}=(I_{0} I_{1}^*+I_{0}^* I_{1})$ and $I_{2}^{\prime}=(I_{0} I_{2}^*+I_{0}^* I_{2})$. Now, We carry out numerical simulations of these theoretical expressions to gain a clear understanding of our experimental results.

\begin{figure*}
\includegraphics[width=\textwidth]{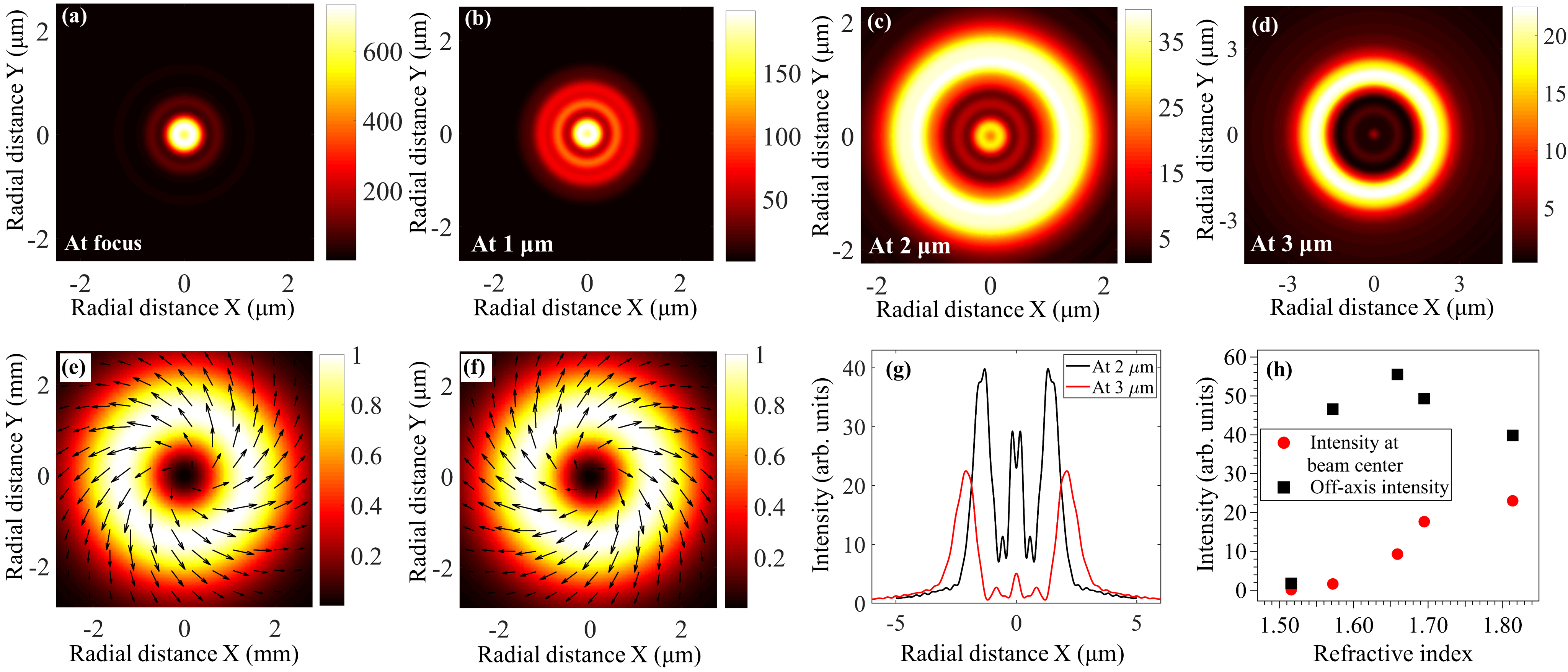}
\caption{  Numerical simulation of the intensity distribution at different positions: (a) at the focus, (b) at \(z = 1 \ \mu m\), (c) at \(z = 2 \ \mu m\), and (d) at \(z = 3 \ \mu m\) away from the focus for the input ACW-ARP (or CW-ARP) vector beam. At the focus, the intensity is high with a small spatial extent. As we move away from the focus, off-axis intensity annular rings form, which are essential for trapping particles off-axis. However, beyond \(z = 2 \ \mu m\), the intensity at the beam center reduces significantly, as shown in (d). (e) and (f) display the intensity and polarization (black arrows) distributions of the incident ACW-ARP (\(\mathbf{E}_{\text{rad}} + \mathbf{E}_{\text{azi}}\)) and CW-ARP (\(\mathbf{E}_{\text{rad}} - \mathbf{E}_{\text{azi}}\)) vector beams across the beam cross-sections, respectively. (g) A line plot of the intensity distribution across the \(x\)-axis at \(z = 2 \ \mu m\) and \(z = 3 \ \mu m\) away from the focus is presented for better visualization. (h) Comparison of the maximum intensity values at the beam center (solid red circles) and at the off-axis annular rings (solid black squares) as a function of the refractive index (RI) at \(z = 2 \ \mu m\) away from the focus.}
\label{Intensity}
\end{figure*}

\section{NUMERICAL SIMULATIONS}
The integral formalism to determine the focused field is utilized in a MATLAB code with parameters from the real experimental setup. As stated earlier, the four-layer stratified medium incorporates a) a microscope immersion oil, b) a coverslip with a refractive index of 1.814, c) a dispersion of sample in deionized water (DI), and d) a top cover glass which is considered semi-infinite, as shown in Fig.~\ref{schematic}(a). The interfaces are positioned at distances of $-165\mu m$ (for the oil-coverslip interface), $160~\mu m$ (for the coverslip-water interface), and $5\mu m$ relative to the nominal focus of the objective lens, which is considered as the origin of the coordinate system. Thus, the layer thicknesses amount to $5~\mu m$ for the oil, $160~\mu m$ for the coverslip, and $35~\mu m$ for the water (which holds LC particles) respectively. With this simulation setup, the intensity distribution is first examined, followed by an investigation of the spin distributions. 

\subsection{Study of the intensity pattern}

In this simulation setup, the first step involves examining the transverse intensity distribution of both ARP vector beams before and after focusing. Both ARP vector beams exhibit a polarization (or intensity) singularity at the beam center before focusing, as shown in Figs.~\ref{Intensity}(e) and (f). The quiver plots over the intensity distribution illustrate the polarization direction of the ARP vector beams. However, after tight focusing of the ARP vector beams, the contributions from the radial and azimuthal components need to be evaluated separately. It has been demonstrated earlier \cite{kumar2024probing,kumar2024inhomogeneous} that while the singularity in a radially polarized beam at the beam center vanishes after tight focusing, an azimuthally polarized beam maintains its singularity even after focusing. Thus, due to the linear nature of the fields, the radial component within both the ARP vector beams consistently contributes to the energy at the center of the focused light, facilitating particle trapping therein. 

This is apparent from Eqs. \ref{electric} and \ref{magnetic}, where the longitudinal component (or the \(z\)-component) for both beams is solely influenced by the \(I_0\) diffraction integrals. In Figs.~\ref{Intensity}(a), (b), (c), and (d), we show the plots of the transverse (xy) intensity distribution at the focus, and at \(1~\mu m\), \(2~\mu m\), and \(3~\mu m\) away from the focus, respectively. These plots reveal the spatial extent of intensity at both on-axis and off-axis positions for the input ARP vector beams. As we move away from the focus, an off-axis annular intensity ring is formed, as shown in Figs.~\ref{Intensity}(c) and (d), which is crucial for observing spatially resolved helicities of light with \(\sigma = +1\) and \(\sigma = -1\). In Fig.~\ref{Intensity}(g), we compare the intensity distributions at \(2~\mu m\) and \(3~\mu m\) away from the focus. This comparison illustrates that, beyond \(2~\mu m\), the intensity at the beam center decreases significantly. Therefore, \(z = 2 \ \mu m\) is the optimized distance from the focus to probe the spatially resolved helicities of light. The increased refractive index (RI) contrast of the second layer within the stratified medium enhances spherical aberration (see Fig.~\ref{schematic}(a)), leading to the formation of these high-intensity rings. The integrals listed in Eq.~\ref{13} offer an intuitive understanding of the field distribution. Among these integrals, \(I_0\), which involves the \(0^{th}\)-order Bessel function \(J_0\), and \(I_1\) and \(I_2\), which involve the \(1^{st}\)-order Bessel function \(J_1\), govern the field distributions at the central and off-axis regions, respectively. A comparative analysis of the highest intensity values in both the central region and the off-axis rings of the intensity profile, across various commercially available coverslips (as shown in Fig.~\ref{Intensity}(h)), indicates that the coverslip used in our experiment, with an RI of 1.814, is well-suited for trapping large particles (size \(\sim 1-10 \ \mu m\) in diameter) at the center and smaller particles (size \(\sim 1-2 \ \mu m\) in diameter) at off-axis positions. This suitability is reflected in the central intensity being approximately \(50-60\%\) of the highest off-axis intensity.

\begin{figure*}
\includegraphics[width=\textwidth]{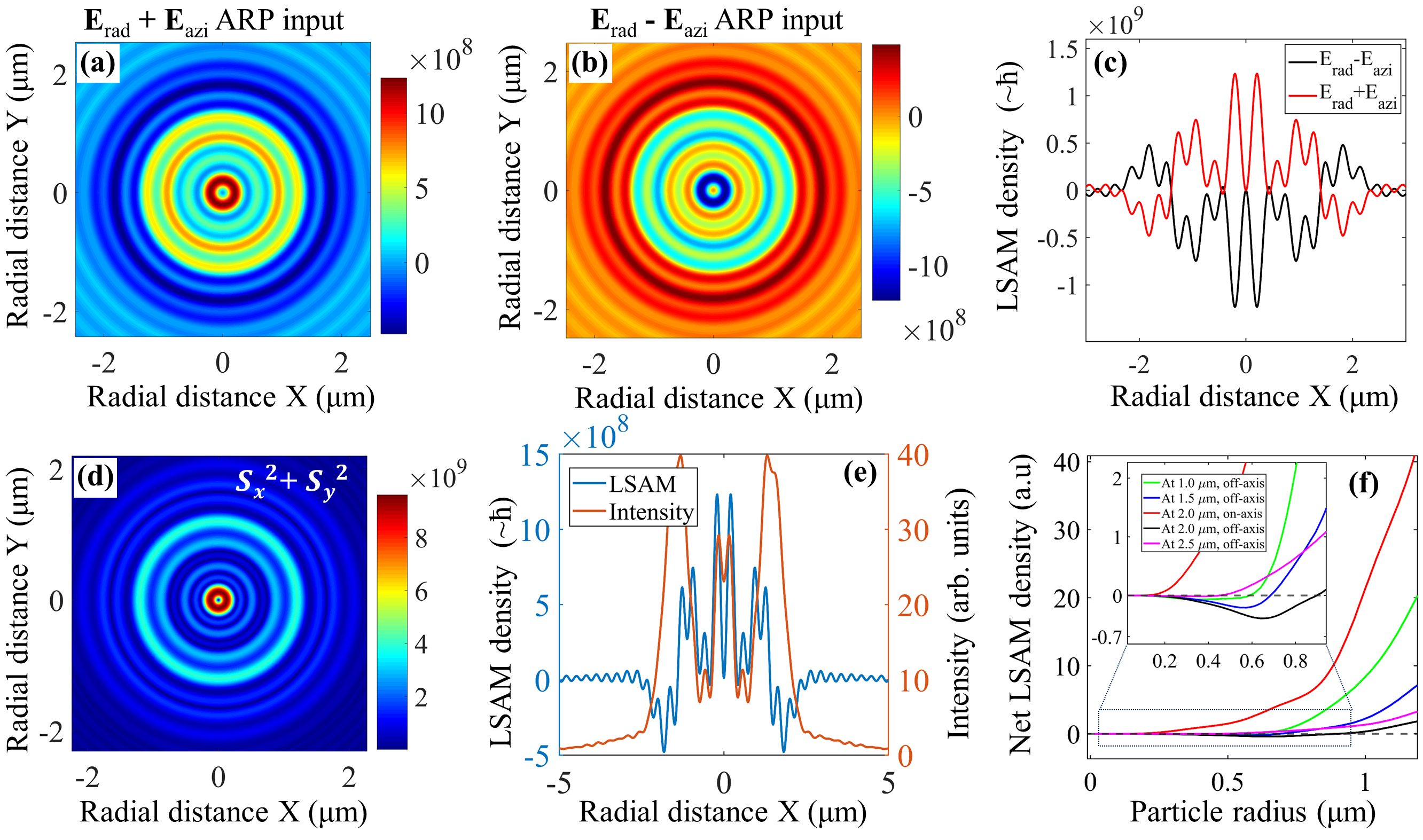}
\caption{Simulated longitudinal spin distribution (LSAM) for the incident (a) ACW-ARP (\(\mathbf{E}_{\text{rad}} + \mathbf{E}_{\text{azi}}\)) and (b) CW-ARP (\(\mathbf{E}_{\text{rad}} - \mathbf{E}_{\text{azi}}\)) vector beams, showcasing the aggregation of LSAM components in concentric circular rings. (c) The line plot of spatially resolved positive/negative (\(\sigma = \pm 1\)) and negative/positive (\(\sigma = \mp 1\)) helicities of the ACW-ARP/CW-ARP vector beams is shown by the red/black line, respectively, with opposite helicities accumulating at different radii. (d) The distributions of the transverse SAM (TSAM) density also show similar aggregation in circular regions. (e) Line plots of LSAM density (light blue) and intensity (orange) at \(z = 2 \ \mu m\) from the focal point, showing the overlap of spatially resolved LSAM and intensity peaks at both on-axis and off-axis positions. (f) The net spatially resolved LSAM (obtained by integrating the local spin density over the area covered by the particle) experienced by a particle as a function of its radius, when placed at the beam center or in the off-axis high-intensity ring, shows a critical radius of \(\sim 1 \ \mu m\) (or diameter \(\sim 2 \ \mu m\)). Below this radius, the particle will experience negative/positive helicity and rotate clockwise/counterclockwise for the incident ACW-ARP/CW-ARP vector beam at off-axis positions. Beyond this radius, particles will always experience positive/negative helicity and rotate counterclockwise/clockwise due to the net positive/negative integration value. These distributions are computed in the transverse plane between \(z = 1 \ \mu m\) and \(z = 2.5 \ \mu m\) beyond the focus, for a coverslip with an RI of 1.814. However, \(z = 2 \ \mu m\) is the optimized distance from the focus for probing the off-axis distribution of LSAM.}
\label{SAM}
\end{figure*}

\subsection{Study of the spin angular momentum density} 

The transverse intensity distributions both off-axis and along the axis suggest similar interesting patterns in other dynamic electromagnetic quantities. Of these, our focus, understandably, would be on investigating the LSAM density - since this is behind the rotation events we report in this work (see Figs.~\ref{SAM} (a)-(f)). It is clear that the LSAM aggregates in concentric rings in the transverse plane, similar to the intensity rings we depicted in Figs.~\ref{Intensity} (c) and (g). In the case of the ACW-ARP vector beam, positive helicity components ($\sigma_{+}$) of the LSAM primarily accumulate in the central rings (Fig.~\ref{SAM}(a)), while negative helicity components ($\sigma_{-}$) accumulate further outward (off-axis). A similar distribution with flipped helicities is observed for the CW-ARP vector beam (Fig.~\ref{SAM}(b)). In Fig.~\ref{SAM}(c)), we demonstrate a line plot that illustrates the range of radii for positive (negative) helicity, extending from the center (radius, $r=0$) to around $1.5\mu m$, while the negative (positive) helicity component spans from around $1.5\mu m$ to $2\mu m$ for an ACW-ARP (CW-ARP) vector beam. In Fig.~\ref{SAM}(d), we present the distributions of TSAM at \(z = 2 \ \mu m\) from the focal point for both ARP vector beams. The TSAM distribution is in the form of concentric rings, similar to the LSAM pattern. As mentioned earlier, subtle focusing and defocusing effects were observed in the experiment as the particle rotated either clockwise or anticlockwise due to the effect of TSAM (see Video 3 in the online Supplementary Information). In Fig.~\ref{SAM}(e), we present the line plots of LSAM density (blue) and intensity (orange) at \(z = 2 \ \mu m\) from the focal point, illustrating the overlap of spatially resolved LSAM and intensity peaks at both on-axis and off-axis positions. Notably, the high-intensity annular ring at the off-axis position overlaps with a negative LSAM density peak for an input ACW-ARP vector beam, ensuring that a particle trapped in this region will spin in the clockwise direction, as observed in our experiments.

Typically the angular momentum experienced by a particle can be assumed to be influenced by the net LSAM within a circular area covered by the particle. Here the net LSAM is the integrated value of the local LSAM density multiplied by the area covered.  To perceive the size dependence of the particle's rotational motion, the net LSAM within a circular region centered at the different locations in the transverse plane is investigated. This net LSAM predicts the direction of the particle's rotation. If it is positive, the particle will rotate counter-clockwise and if negative, the particle will rotate clockwise. The radii of the circular areas are increased gradually to consider particles of diverse sizes trapped at on-axis and off-axis spatial locations across the transverse plane. Thus, two particles -- one centered at the trap center (simulating an on-axis trap) and another off-center (simulating an off-axis trap) -- exhibit distinctly different spin dynamics. A particle trapped at the beam center will rotate counter-clockwise/clockwise under the influence of the incident ACW-ARP/CW-ARP vector beam, regardless of its size, since the SAM at the center is much higher in value compared to that off-axis (see Fig.~\ref{schematic} (a)). Conversely, a particle trapped in a high-intensity off-axis ring will rotate in a direction determined by its size. Small particles, with a radius around 0.5-1~$\mu$m, will rotate clockwise/anti-clockwise for the ACW-ARP/CW-ARP vector beam, while larger particles of radius greater than $2~\mu$ m will rotate counterclockwise / clockwise.

In Fig.~\ref{SAM}(f), we show the results for an input ACW-ARP vector beam. The reverse pattern can be observed for the CW-ARP vector beam (not depicted here). Thus, the critical radius of the particle required to demonstrate a transition from CW/ACW to ACW/CW rotation at a trapping plane $z = 2\mu m$ away from the focus is numerically found to be $1 \mu m$, matching very well with the experiment. Also, the value of the off-axis net LSAM is much lesser than the on-axis net LSAM, which explains the lower frequency of rotation for small particles trapped at the off-axis intensity ring as observed in the experiment (see Video 5 in the online Supplementary Information). Note that the optimized distance from the focus is \( z = 2 \ \mu\text{m} \), where the net LSAM density at off-axis positions exhibits the extremum of opposite helicity (negative/positive value of LSAM for ACW-ARP/CW-ARP input beam). Before and after \( z = 2 \ \mu\text{m} \), the off-axis net LSAM density decreases. As a result, the critical radius at which rotational reversal occurs may vary depending on the depth of trapping and the RI of the coverslips used in the experiment. Changes in RI affect spherical aberration, which in turn influences both the intensity and LSAM ring diameters in the focused light.     

\section{Conclusion}
In conclusion, the SOI of light due to the tight focusing of structured ARP vector beams is investigated both numerically and experimentally and is utilized in optical tweezers to engineer the spin dynamics of birefringent microparticles at different spatial locations. Due to the tight focusing of these beams having a net zero angular momentum, the different components of spin angular momentum are observed to be spatially resolved into concentric regions in the transverse plane. This form of the Spin-Hall effect is influenced by the underlying axial symmetry of the electric fields across the beam cross-section. Experiments with spin-responsive liquid crystal (LC) microspheres in optical tweezers, when illuminated by these beams, reveal distinct signatures of spin dynamics induced by opposite helicities at different spatial locations. Particles trapped along the beam propagation axis will always rotate following the direction of the electric field's rotation (or spiral polarization) across the beam section, rotating clockwise for the CW-ARP vector beam and counterclockwise for the ACW-ARP vector beam.
Conversely, particles trapped at off-axis locations exhibit size-dependent directional rotation, with small particles rotating counter to the rotation of the electric field in the beam cross-section, while larger particles will follow the rotation of the electric field. Most importantly, the smaller particle can be made to spin in opposite directions by trapping it either on-axis or off-axis by changing the focus of the trapping objective lens. Consequently, this work presents an experimentally viable strategy for engineering optical traps capable of inducing controlled and specific spin motions of trapped particles using a single trapping spinless ARP vector beam. Importantly,  these effects are a beautiful manifestation of the diverse possibilities offered by spin-orbit optomechanics, that precludes the need to structure complex beam profiles using advanced algorithms involving adaptive optics for manipulation of particles in optical tweezers. In the future, we intend to demonstrate a new class of optical micromachines driven by spin-orbit optomechanics using similar mechanisms.

The authors acknowledge the SERB, Department of Science and Technology, Government of India (Project No. EMR/2017/001456) and IISER Kolkata IPh.D fellowship for research.



\providecommand{\noopsort}[1]{}\providecommand{\singleletter}[1]{#1}%

\end{document}